\documentstyle[12pt]{article}

\makeatletter
\def\vereq#1#2{\lower3pt\vbox{\baselineskip1.5pt \lineskip1.5pt
\ialign{$\m@th#1\hfill##\hfil$\crcr#2\crcr\sim\crcr}}}
\makeatother

\def\lesssim{\mathrel{\mathpalette\vereq<}}
\def\gtrsim{\mathrel{\mathpalette\vereq>}}

\makeatletter
\def\lesssim{\mathrel{\mathpalette\vereq<}}
\def\vereq#1#2{\lower3pt\vbox{\baselineskip1.5pt \lineskip1.5pt
\ialign{$\m@th#1\hfill##\hfil$\crcr#2\crcr\sim\crcr}}}

\def\gtrsim{\mathrel{\mathpalette\vereq>}}
\makeatother

\newcommand{\beq}{\begin{equation}}
\newcommand{\eeq}{\end{equation}}

\newcommand{\remove}[1]{}

\begin{document}
\begin{titlepage}
\begin{center}
\today     \hfill    LBL-38380 \\
~{} \hfill UCB-PTH-96/06  \\

\vskip .25in

{\large \bf A Supersymmetric Theory of 
Flavor and $R$ Parity}\footnote{This 
work was supported in part by the Director, Office of 
Energy Research, Office of High Energy and Nuclear Physics, Division of 
High Energy Physics of the U.S. Department of Energy under Contract 
DE-AC03-76SF00098 and in part by the National Science Foundation under 
grant PHY-90-21139.}

%

\vskip 0.3in

Christopher D. Carone$^1$, Lawrence J. Hall$^{1,2}$, and 
Hitoshi Murayama$^{1,2}$

\vskip 0.1in

{{}$^1$ \em Theoretical Physics Group\\
     Lawrence Berkeley National Laboratory\\
     University of California, Berkeley, California 94720}

\vskip 0.1in

{{}$^2$ \em Department of Physics\\
     University of California, Berkeley, California 94720}

\end{center}

\vskip .3in

\begin{abstract}
We construct a renormalizable, supersymmetric theory of flavor 
and $R$ parity based on the discrete flavor group $(S_3)^3$.
The model can account for all the masses and mixing angles of
the Standard Model, while maintaining sufficient squark degeneracy
to circumvent the supersymmetric flavor problem.  By starting with
a simpler set of flavor symmetry breaking fields than we have
suggested previously, we construct an economical Froggatt-Nielsen 
sector that generates the desired elements of the fermion Yukawa 
matrices.  With the particle content above the flavor scale completely
specified, we show that all renormalizable $R$-parity-violating interactions
involving the ordinary matter fields are forbidden by the flavor symmetry. 
Thus, $R$ parity arises as an accidental symmetry in our model.
Planck-suppressed operators that violate $R$ parity, if present, can be 
rendered harmless by taking the flavor scale 
to be $\lesssim 8 \times 10^{10}$ GeV.
\end{abstract}

\end{titlepage}
\renewcommand{\thepage}{\roman{page}}
\setcounter{page}{2}
\mbox{ }

\vskip 1in

\begin{center}
{\bf Disclaimer}
\end{center}

\vskip .2in

\begin{scriptsize}
\begin{quotation}
This document was prepared as an account of work sponsored by the United
States Government. While this document is believed to contain correct 
information, neither the United States Government nor any agency
thereof, nor The Regents of the University of California, nor any of their
employees, makes any warranty, express or implied, or assumes any legal
liability or responsibility for the accuracy, completeness, or usefulness
of any information, apparatus, product, or process disclosed, or represents
that its use would not infringe privately owned rights.  Reference herein
to any specific commercial products process, or service by its trade name,
trademark, manufacturer, or otherwise, does not necessarily constitute or
imply its endorsement, recommendation, or favoring by the United States
Government or any agency thereof, or The Regents of the University of
California.  The views and opinions of authors expressed herein do not
necessarily state or reflect those of the United States Government or any
agency thereof, or The Regents of the University of California.
\end{quotation}
\end{scriptsize}

\vskip 2in

\begin{center}
\begin{small}
{\it Lawrence Berkeley Laboratory is an equal opportunity employer.}
\end{small}
\end{center}

\newpage
\renewcommand{\thepage}{\arabic{page}}
\setcounter{page}{1}
\section{Introduction}

In supersymmetric models of particle physics there are two aspects to the
flavor problem. The first is the problem of quark and lepton mass and mixing
hierarchies: why are there a set of small dimensionless Yukawa couplings in
the theory? The second aspect of the problem is why the superpartner gauge
interactions do not violate flavor at too large a rate. This requires that the
squark and slepton mass matrices not be arbitrary.  Rather, these matrices
must also possess a set of small parameters which suppresses flavor-changing
effects, even though all the eigenvalues are large. What is the origin of
this second set of small dimensionless parameters? 

An extremely attractive hypothesis is to assume that the two sets of small
parameters, those in the fermion mass matrices and those in the scalar mass
matrices, have a common origin: they are the small symmetry breaking
parameters of an approximate flavor symmetry group $G_f$. This provides a link
between the fermion mass and flavor-changing problems; both are addressed by
the same symmetry. Such an approach was first advocated using a flavor group
$U(3)^5$, broken only by the three Yukawa matrices $\lambda_{U,D,E}$ in the
up, down and lepton sectors \cite{HR}. This not only solved the
flavor-changing problem, but suggested a boundary condition on the soft
operators which has a more secure theoretical foundation than that of
universality. However, this framework did not provide a model for the origin
of the Yukawa matrices themselves, and left open the possibility that $G_f$
was more economical than the maximal flavor group allowed by the standard
model gauge interactions. 

The first explicit models in which spontaneously broken flavor groups were
used to constrain both fermion and scalar mass matrices were based on $G_f =
SU(2)$ \cite{DKL} and $G_f = U(1)^3$ \cite{NS}. In the first case the
approximate degeneracy of scalars of the first two generations was guaranteed
by $SU(2)$ in the symmetry limit.  In retrospect it seems astonishing that the
flavor-changing problem of supersymmetry was not solved by such a flavor group
earlier. The well known supersymmetric contributions to the $K_L$-$K_S$ mass
difference can be rendered harmless by making the $\tilde{d}$ and $\tilde{s}$
squarks degenerate \cite{DG}. Why not guarantee this degeneracy by placing 
these squarks in a doublet of a non-Abelian flavor 
group $(\tilde{d}, \tilde{s})$? In the case of Abelian $G_f$, the squarks 
are far from degenerate, however it was discovered that the flavor-changing 
problem could be solved by arranging for the Kobayashi-Maskawa mixing 
matrix to have an origin in the up sector rather than the down sector. 

A variety of supersymmetric theories of flavor have followed, including ones
based on $G_f =O(2)$ \cite{PS}, $G_f=U(1)^3$ \cite{LNS}, $G_f=\Delta (75)$
\cite{KS}, $G_f=(S_3)^3$ \cite{HM,CHM} and $G_f = U(2)$ \cite{BDH}. Progress
has also been made on relating the small parameters of fermion and scalar mass
matrices using a gauged $U(1)$ flavor symmetry in a $N=1$ supergravity theory,
taken as the low energy limit of superstring models \cite{DPS}. Development of
these and other theories of flavor is of great interest because they offer the
hope that an understanding of the quark and lepton masses, and the masses of
their scalar superpartners, may be obtained at scales well beneath the Planck
scale, using simple arguments about fundamental symmetries and how they are
broken. The theories, to varying degrees, give understanding to the patterns
of the mass matrices, and may, in certain cases, also lead to very definite
mass predictions. Furthermore, flavor symmetries may be of use to understand a
variety of other important aspects of the theory. 

The general class of theories which address both aspects of the supersymmetric
flavor problem have two crucial ingredients: the flavor group $G_f$ and the
flavon fields $F$, which have a hierarchical set of vacuum expectation
values (vevs) allowing a sequential breaking of $G_f$\footnote{We assume that
the scalar mass squared matrices are constrained by the flavor symmetry,
i.e., that the messenger scale of supersymmetry breaking is higher
than the flavor scale.}. These theories can be specified in two very 
different forms. In the first form, the only fields in the theory
beyond $F$ are the light matter and Higgs fields. An effective theory is
constructed in which all gauge and $G_f$ invariant interactions are written
down, including non-renormalizable operators scaled by some mass scale of
flavor physics, $M_f$. The power of this approach is that considerable
progress is apparently possible without having to make detailed assumptions
about the physics at the scale $M_f$ which generates the non-renormalizable
operators. Much, if not all, of the flavor structure of fermion and scalar
masses comes from such non-renormalizable interactions, and it is interesting
to study how their form depends only on the choice of $G_f$, how $G_f$ is
broken, and the light field content. 

A second, more ambitious, approach is to write a complete, renormalizable
theory of flavor at the scale $M_f$. Such a theory possesses a set of heavy
fields which, when integrated out of the theory, lead to the effective theory
discussed above \cite{FN}. However, it is reasonable to question whether the
effort required to construct such full theories is warranted. Clearly these
complete theories involve further assumptions beyond those of the effective
theories, namely the $G_f$ properties of the fields of mass $M_f$, and it
would seem that the low energy physics of flavor is independent of this,
depending only on the properties of the effective theory. In
non-supersymmetric theories such a criticism may have some validity, but in
supersymmetric theories it does not. This is because in supersymmetric
theories, on integrating out the states of mass $M_f$, the low energy theory
is not the most general effective theory based on the flavor group 
$G_f$. Several operators which are $G_f$ invariant, and could 
be present in the effective theory, are typically not generated when the 
heavy states of mass $M_f$ are integrated out. Which operators are missing 
depends on what the complete theory at $G_f$ looks like. This phenomena is well
known, and is illustrated, for example, in 
references \cite{LNSF,ADHRS,KS,BDH}, and it casts doubt on the effective 
theory approach to building supersymmetric theories of flavor. Finally, 
one might hope that a complete renormalizable theory of flavor at 
scale $M_f$ might possess a simplicity which is partly hidden at the 
level of the effective theory. 

We have previously discussed an effective theory of flavor based on the 
gauged flavor group $G_f =(S_3)^3$ \cite{HM,CHM}. In this paper we find a 
simple, complete, renormalizable theory with $G_f=(S_3)^3$, and we 
demonstrate, that acceptable fermion and scalar mass matrices result 
from integrating out the heavy states.  In addition, we discover an origin 
for $R$ parity in the $G_f$ properties of the renormalizable interactions of 
the complete theory. In the effective theory approach there 
are $R$-parity-violating operators which are $G_f$
allowed and must be forbidden by hand to avoid phenomenological difficulties.
However, such operators are not generated from our full theory: we can
understand $R$ parity to be an unavoidable consequence of the $G_f$ structure
of the Higgs and matter representations of the complete theory. 

Our choice of a gauged $(S_3)^3$ as the flavor group is motivated by a 
number of considerations.  First, we choose a gauged flavor symmetry over 
a global one to avoid the criticism that global symmetries 
are not respected by quantum gravitational effects.  If
the gauged flavor symmetry is a continuous one \cite{DKL}, then there will be
D-term contributions to the scalar potential that couple ordinary
squarks to the flavon fields.  In this case, flavon expectation values 
may generate substantial nonuniversal contributions to the squark 
masses, and hence, dangerous flavor changing neutral current effects
\cite{NMR}. We therefore choose to work with a discrete gauged flavor
symmetry, for which there are no associated $D$-terms.  We then choose a
discrete group that has both ${\bf 2}$ and ${\bf 1}$ dimensional 
representations.  With this representation structure, we can embed the 
chiral superfields of the first two generations into the doublet,
to maintain the near degeneracy of the corresponding squarks.
The smallest discrete flavor group with these representations
is $S_3$, which has a ${\bf 2}$, ${\bf 1}_S$, and ${\bf 1}_A$. 
The latter is a one-dimensional representation that transforms
nontrivially under the group.  We assign the third generation fields
to the ${\bf 1}_A$ rather than ${\bf 1}_S$ so that the model
is free of discrete gauge anomalies.  The three generations of the 
standard model therefore correspond to the representation structure
${\bf 2}$+${\bf 1}_A$.  If we tried to build a model in which $G_f$ 
involved only a single $S_3$ factor, we would find that it is impossible 
to explain the hierarchy between, for example, the down and strange quark 
masses, which both would be invariant under the flavor group.  A simple
way around this problem is to replicate $S_3$ factors, so that
the left-handed doublet fields $Q$, and the right-handed
singlet fields $U$ and $D$ each transform under a different $S_3$.
In addition, if the Higgs fields are chosen to transform as ${\bf 1}_A$'s
under both $S_3^Q$ and $S_3^U$ simultaneously, only the top quark Yukawa
coupling is left invariant under the flavor symmetry.  The remaining quark
Yukawa couplings can be treated as small symmetry-breaking spurions, 
and the deviation from squark degeneracy easily estimated.  This analysis
was carried out in Ref.~\cite{HM}, where it was shown that the forms of the 
squark mass-squared matrices were phenomenologically viable.
In addition, the model can be extended to the lepton sector by assigning 
the doublet chiral superfield $L$ and the singlet $E$ to 
${\bf 2}+{\bf 1}_A$'s of $S_3^D$ and $S_3^Q$, respectively~\cite{CHM}.  
This leads to acceptable slepton mass-squared matrices and a distinctive 
proton decay signature that may be within the reach of
SuperKamiokande~\cite{CHM}.  

It is the point of our current work to explain how an acceptable pattern 
of $(S_3)^3$ breaking originates at a fundamental level, and to show
how $R$ parity emerges from the flavor structure of the full theory.
Unlike Refs.~\cite{HM,CHM}, we will allow the flavor scale $M_f$ to be
considerably lower than the Planck scale $M_{Pl}$. In this case, 
the constraints from proton decay on the acceptable flavon
quantum number assignments \cite{CHM} are considerably weakened.
This in turn allows us to construct a much more elegant model.  The paper is
organized as follows.  In the Section~\ref{sec:supp} we review the known
mechanisms of suppressing baryon- and lepton-number-violating interactions in 
supersymmetric models. In Section~\ref{sec:model}, we present the quantum
number assignments for the flavor symmetry breaking fields $F$ in our model. 
We show that the most general set of higher dimension operators involving the
$F$ fields generate viable fermion Yukawa matrices when the flavons acquire
vevs.  In addition, we show that the pattern of flavor symmetry breaking in 
our model leads to squark and slepton mass-squared matrices that are
phenomenologically acceptable.  In Section~\ref{sec:froggatt}, we 
present a renormalizable model that generates the necessary operators 
involving the $F$ fields when a set of vector-like fields are integrated 
out beneath the flavor scale $M_f$.  Given the field content above the
scale $M_f$, we show that all renormalizable $R$-parity-violating
operators are forbidden by the flavor symmetry.  We also take into
account the possibility of nonrenormalizable $R$-parity-violating operators
generated at the Planck scale.  In the final section, we summarize our
conclusions.  In an appendix we provide an example of 
a workable potential that generates the pattern of vevs assumed in the 
main body of the paper.

\section{The suppression of baryon and lepton number violation.} 
\label{sec:supp}

The standard model, for all its shortcomings, does provide an understanding
for the absence of baryon ($B$) and lepton ($L$) number violation: the field
content simply does not allow any renormalizable interactions which violate
these symmetries. This is no longer true when the field content is extended to
become supersymmetric; squark and slepton exchange mediate baryon and lepton
number violation at unacceptable rates, unless an extra symmetry, such as $R$
parity, is imposed on the theory.  The need for a new symmetry, which in
general we label $X$,  was first realised in the context of a supersymmetric
$SU(5)$ grand unified theory \cite{WI}. As will become clear, there are a wide
variety of possibilities for the $X$ symmetry. Matter 
parity\footnote{Matter parity is equivalent to $R$ parity, up to a
$2\pi$ rotation.} \cite{DG}, $Z_N$ symmetries other than matter 
parity \cite{HS,BHR,IR} and baryon or lepton numbers \cite{DH,CM,CM2} 
provide well known examples, each giving a distinctive phenomenology. One 
of the most fundamental questions in constructing supersymmetric models 
is \cite{W,SY} {\it What is the origin of this extra symmetry needed 
to suppress baryon and lepton number violating processes?} 

The $X$ symmetry must have its origin in one of the three categories of
symmetries which occur in field theory models of particle physics: spacetime
symmetries, gauge (or vertical) symmetries and flavor (or horizontal)
symmetries. The $X$ symmetry is most frequently referred to as 
$R$ parity\footnote{$R_p$ was first introduced in a completely different 
context \cite{FF}.}, $R_p$, which is a $Z_2$ parity acting on the 
anti-commuting coordinate of superspace and on the chiral superfields,
such that $\theta \to - \theta$, matter fields$\to -$matter fields and 
higgs fields $\to $higgs fields.  We view this as unfortunate, since 
it suggests that the reason for the suppression of baryon and lepton
number violation is to be found in spacetime symmetries, which certainly need
not be the case. $R_p$ can be viewed as a superspace analogue of the familiar
discrete spacetime symmetries, such as $P$ and $CP$. In the case of $P$ and
$CP$ we know that they can appear as accidental symmetries in gauge models
which are sufficiently simple. For example $P$ is an accidental symmetry of
QED and QCD, while CP is an accidental symmetry of the two generation standard
model. Nevertheless, in the real world $P$ and $CP$ are broken. This suggests
to us that discrete spacetime symmetries are not fundamental and should not be
imposed on a theory, so that if $R_p$ is a good symmetry, it should be
understood as being an accidental symmetry resulting from some other symmetry.
These arguments can also be applied to alternative spacetime origins for $X$,
such as a $Z_4$ symmetry on the coordinate $\theta$ \cite{HS}. \footnote{
Clearly these arguments need not be correct: for example, it could be that
both $P$ and $CP$ are fundamental symmetries, but they have both been
spontaneously broken. However, in this case the analogy would suggest that
$R_p$ is also likely to be spontaneously broken.} Hence, while the symmetry
$X$ could have a spacetime origin, we find it more plausible that it arises
from gauge or flavor symmetries. 

In this case what should we make of $R_p$? If it is a symmetry at all, it
would be an accidental symmetry, either exact or approximate. If $R_p$ is
broken by operators of dimension 3, 4 or 5, then a weak-scale, lightest
superpartner (LSP) would not be the astrophysical dark matter. The form of the
$R_p$ breaking interactions will determine whether the LSP will decay in
particle detectors or whether it will escape leaving a missing energy
signature. The realization that $X$ may well have an origin in gauge or flavor
symmetries, has decoupled the two issues of the suppression of $B$ and $L$
violation, due to $X$, and the lifetime of the LSP, governed by $R_p$
\cite{BHR,H}. 

At first sight, the most appealing origin for $X$ is an extension of the
standard model gauge group, either at the weak scale \cite{W}, or at the grand
unified scale \cite{SY}. An interesting example is provided by the crucial
observation that adding $U(1)_{B-L}$ \cite{SY}, or equivalently
$U(1)_{T_{3R}}$, is sufficient to remove all renormalizable $B$ and $L$
violation from the low energy theory: matter parity is a discrete subgroup of
$U(1)_{B-L}$. This is clearly seen in SO(10) \cite{CHH},
where the requirement that all interactions have an even number of spinor
representations immediately leads to matter parity. 

However, this example has a gauge group with  rank larger than that of the
standard model, and the simplest way to spontaneously reduce the rank, for
example via the vev of a spinor {\bf 16}-plet in $SO(10)$, leads to a large
spontaneous breaking of the discrete matter parity subgroup of $SO(10)$
\cite{M,KM}. Thus theories based on $SO(10)$ need a further ingredient to
ensure sufficient suppression of $B$ and $L$ violation of the low energy 
theory.  One possibility is that the spinor vev does not introduce the 
dangerous couplings, which typically requires a discrete symmetry 
beyond $SO(10)$. Alternatively the rank may be broken by a larger Higgs 
multiplets \cite{M}, for example the {\bf 126} representation of $SO(10)$. 
Finally, if the reduction of rank occurs at low energies, the 
resulting $R_p$ violating phenomenology may be acceptable \cite{KM}, however, 
the weak mixing angle prediction is then lost (For exceptions,
see Refs.~\cite{chenghall}). The flipped $SU(5)$ gauge group allows for 
models with renormalizable $L$ violation, but highly suppressed $B$ 
violation \cite{BrH}; however, these theories also lose the weak mixing 
angle prediction. 

There are other possibilities for $X$ to be a discrete subgroup of an enlarged
gauge symmetry. Several $Z_N$ examples from $E_6$ are possible \cite{BHR}.
Such a symmetry will be an anomaly free discrete gauge symmetry, and it has
been argued that if $X$ is discrete it should be anomaly free in order not to
be violated by Planck scale physics \cite{KW}. With the minimal low energy
field content, there are only two such possibilities which commute with
flavor: the familiar case of matter parity, and a $Z_3$ baryon parity
\cite{IR}, which also prohibits baryon number violation from dimension 5
operators. While the gauge origin of $X$ remains a likely possibility, we are
not aware of explicit compelling models which achieve this. 

Another possible mechanism of suppressing $R$-parity violation, which
is not discussed in the literature, is a Peccei--Quinn symmetry.  This
anomalous global symmetry was proposed in Ref.~\cite{PQ} to solve the
strong CP problem in QCD.  In the context of supersymmetric models, we
assign the same charge $+1$ to all the matter chiral superfields, $Q$, $U$,
$D$, $L$, and $E$, and a charge $-2$ to the Higgs chiral superfields
$H_u$ and $H_d$.  This symmetry forbids all $R$-parity violating
interactions.  If we break the Peccei--Quinn symmetry using a field with
even charges, it leaves an unbroken $Z_2$ symmetry which is nothing but
the matter parity that we have discussed.  The same Peccei--Quinn
symmetry forbids the $B$-violating dimension-five operators in the
symmetry limit, but they are induced by its breaking in general.  The
extent of suppression depends on the details of the models
\cite{CRHRRV,HMY,HMTY}. 

Finally we discuss the possibility that the $X$ symmetry is a flavor symmetry:
the symmetry which is ultimately responsible for the small parameters of the
quark and lepton mass matrices, and also of the squark and slepton mass
matrices, might provide sufficient suppression for $B$ and $L$ violation.
Indeed, this is an extremely plausible solution for the suppression of $L$
violation since the experimental constraints on the coefficients of the $L$
violating interactions are quite weak, and would be satisfied by having
amplitudes suppressed by powers of small lepton masses. However, the
experimental constraints involving $B$ violation are so strong, that
suppression by small quark mass factors are insufficient \cite{HK}. Hence the
real challenge for these theories is to understand the suppression of $B$
violation. 

Some of the earliest models involving matter parity violation had a discrete
spacetime \cite{HS} or gauge \cite{BrH} origin for $B$ conservation, but had
$L$ violation at a rate governed by the small fermion masses. This distinction
between $B$ and $L$ arises because left-handed leptons and Higgs doublets are
not distinguished by the standard model gauge group, whereas quarks are
clearly distinguished by their color. This provides a considerable motivation
to search for supersymmetric theories with matter parity broken only by the
$L$ violating interactions. 

It is not difficult to understand how flavor symmetries could lead to exact
matter parity. Consider a supersymmetric theory, with minimal field content
and gauge group, which has the flavor group $U(3)^5$ broken only by parameters
which transform like the usual three Yukawa coupling matrices. The Yukawa
couplings and soft interactions of the most general such effective theory can
be written as a power series in these breaking parameters, leading to a theory
known as weak scale effective supersymmetry \cite{HR}. The flavor group and
transformation properties of the breaking parameters are sufficient to forbid
matter parity violating interactions to all orders: each breaking parameter
has an even number of $U(3)$ tensor indices, guaranteeing that all
interactions must have an even number of matter fields.\footnote{ This point
was missed in \cite{HR} where $R_p$ was imposed unnecessarily as an additional
assumption. We believe that the automatic conservation of $R_p$ makes this
scheme an even more attractive framework as a model independent low energy
effective theory of supersymmetry.} To construct an explicit model along these
lines it is perhaps simplest to start with a $U(3)$ flavor group, with all
quarks and leptons transforming as triplets, but Higgs doublets as trivial
singlets. An exact matter parity will result if the spontaneous breaking of
this flavor group occurs only via fields with an even triality. A similar idea
has recently been used in the construction of a four generation theory with
gauged flavor $SU(4)$ symmetry \cite{BN,CM2}. 

In view of the recent activity in constructing explicit supersymmetric
theories of flavor \cite{DKL,NS,PS,LNS,KS,HM,CHM,BDH}, an interesting question
is whether the $X$ symmetry is contained in a flavor group \cite{KaM}. With
Abelian flavor groups, the suppression of $L$ violation is quite natural
\cite{BGNN}, while sufficient suppression of $B$ violation is much 
harder to obtain \cite{BHN}. In this paper we construct a theory of flavor
based on the non-Abelian discrete group $(S_3)^3$. It is found to provide an
explanation for the suppression of $B$ and $L$ violation that is analogous to
the matter parity found in $SO(10)$ theories, with the difference, however,
that $B$ and $L$ are not exact. 

\section{The Model}\label{sec:model}

As we described earlier, the three generations of $Q$, $U$, and $D$
fields transform as ${\bf 2}$+${\bf 1}_A$'s under the corresponding
$S_3$ group. The ordinary Higgs fields transform 
as $({\bf 1}_A,{\bf 1}_A,{\bf 1}_S)$'s under $S_3^Q \times
S_3^U \times S_3^D$ .   Given these assignments, the quark Yukawa
matrices have well defined transformation properties under
$(S_3)^3$:
\begin{equation}
Y_u \sim \left(
\begin{array}{cc|c} 
\multicolumn{2}{c|}{
({\bf \tilde{2}},{\bf \tilde{2}},{\bf 1}_S) }
& ({\bf \tilde{2}},{\bf 1}_S,{\bf 1}_S) \\ \hline
\multicolumn{2}{c|}{({\bf 1}_S,{\bf \tilde{2}},{\bf 1}_S)} &
({\bf 1}_S,{\bf 1}_S,{\bf 1}_S) \end{array} \right) 
\,\,\, , \,\,\,
Y_d \sim \left(\begin{array}{cc|c} 
\multicolumn{2}{c|}{
({\bf \tilde{2}},{\bf 1}_A,{\bf 2})}
& ({\bf \tilde{2}},{\bf 1}_A,{\bf 1}_A) \\ \hline
\multicolumn{2}{c|}{({\bf 1}_S,{\bf 1}_A,{\bf 2})} &
({\bf 1}_S,{\bf 1}_A,{\bf 1}_A) \end{array}\right) 
\label{eq:transp}
\end{equation}
where we use the notation ${\bf \tilde{2}} \equiv {\bf 2} \otimes {\bf
1}_A$,\footnote{${\bf \tilde{2}} = (a,b)$ is equivalent to ${\bf 2} =
(b, -a)$.}.  In the lepton sector, the fields $L$ and $E$ transform
in the same way as $D$ and $Q$ under the flavor symmetry, so that the 
lepton Yukawa matrix transforms in the same way as $Y_d^T$. 

We first specify the quantum number assignments for the fields that 
acquire flavor symmetry breaking vevs.  Products of these fields must
have the proper transformation properties to generate (at least some of)
the various blocks of the fermion Yukawa matrices shown in
eq.~(\ref{eq:transp}).  The flavon fields $F$ in our model are
\[ 
\Phi_Q^{(i)} \sim ({\bf 2},{\bf 1}_A,{\bf 1}_S)  \,\,,\,\,\,\,
\Phi_D^{(i)} \sim ({\bf 1}_A,{\bf 1}_S,{\bf 2})  \,\,,\,\,\,\,  
\Phi_U^{(i)} \sim ({\bf 1}_A,{\bf 2},{\bf 1}_S)  \,\,,
\]\begin{equation}
\chi_1 \sim ({\bf 1}_S,{\bf 1}_A,{\bf 1}_A) \,\,,\,\,\,\,  \chi_2 \sim ({\bf
1}_A,{\bf 1}_S,{\bf 1}_A)\,\,,
\label{eq:flavons}
\end{equation}
where $i=1,2$.  Note that these are simpler representations for
the flavon fields than those presented in Refs.~\cite{HM,CHM}.  While 
we argued in Ref.~\cite{CHM} that some of the flavon representations 
shown above were excluded by their contribution to proton decay via 
Planck-suppressed dimension-five operators, we will see in 
Section~\ref{sec:froggatt} that these operators are easily suppressed
by taking the flavor scale to be somewhat below $M_{Pl}$.

Let us now explicitly construct the fermion Yukawa matrices that
follow from (\ref{eq:flavons}).  The two-by-two down-strange and 
up-charm Yukawa matrices involve products of the form 
\begin{equation}
\Phi_Q^{(i)} \Phi_D^{(j)} \sim ({\bf \tilde{2}},{\bf 1}_A,{\bf 2})
\,\,\,\,\,\mbox{and}
\,\,\,\,\, \Phi_Q^{(i)} \Phi_U^{(j)} \sim ({\bf \tilde{2}},{\bf
\tilde{2}},{\bf 1}_S)  
\,\, .
\label{eq:tbtms}
\end{equation}
Each of the eight combinations of $\Phi$ fields shown above
can form a flavor-invariant dimension-six operator that
contributes to the usual Yukawa coupling matrices when the
flavon fields acquire vevs.  For example, the down-strange
block originates from the operators 
\begin{equation}
\frac{1}{M^2_f} \sum_{ij} c^d_{ij} Q H_d \Phi_Q^{(i)} \Phi_D^{(j)} D
\label{eq:dimen6}
\end{equation}
where $M_f$ is the flavor-physics scale, and the $c^d_{ij}$ are
order one coefficients.  Note that we have introduced two $\Phi_Q$ 
doublets in order to assure a nonvanishing Cabibbo angle. In addition, 
we require two $\Phi_U$ and $\Phi_D$ fields so that the up and down quark
masses are both nonvanishing.  This would not be possible if the
Yukawa matrices in (\ref{eq:tbtms}) were each formed from the 
product of exactly two doublets; any matrix constructed in this way 
has a vanishing determinant.  In our discussion below, we will let 
each $\Phi_a$ field (with $a=Q$, $U$, or $D$) represent some linear 
combination of $\Phi_a^{(1)}$ and $\Phi_a^{(2)}$, leaving it implicit 
that different occurrences of $\Phi_a$ may indicate different 
linear combinations.

Let us denote the ratio of the vevs of the $\Phi$ and $\chi$ fields to 
the flavor-physics scale $M_f$ by the parameters $\epsilon$ and $\delta$. 
If we choose the $\Phi$ field vevs 
\beq
\frac{1}{M_f} \langle \Phi_Q \rangle \sim 
\epsilon_Q \left[\begin{array}{c} \lambda  \\ 1 \end{array}
\right] 
\,\,\,\,\,\,\,\, 
\frac{1}{M_f} \langle \Phi_D \rangle \sim 
\epsilon_D \left[\begin{array}{c} \lambda  \\ 1 \end{array}
\right] 
\,\,\,\,\,\,\,\,
\frac{1}{M_f} \langle \Phi_U \rangle \sim 
\epsilon_U \left[\begin{array}{c} \lambda^3 \\ 1 \end{array}
\right]
\eeq
then the down-strange and up-charm Yukawa matrices will take the form
\beq
\epsilon_Q \epsilon_D \left[\begin{array}{cc} \lambda^2 & \lambda \\
                                              \lambda   &   1 
                             \end{array}\right]
\,\,\,\mbox{ and } \,\,\,
\epsilon_Q \epsilon_U \left[\begin{array}{cc} \lambda^4   & \lambda \\
                                              \lambda^3   &    1
                             \end{array}\right]
\label{eq:tbtm}
\eeq
respectively, where $\lambda \approx 0.22$ is the Cabibbo angle.
We set $\epsilon_Q\epsilon_D \sim \lambda^5$ and 
$\epsilon_Q\epsilon_U \sim \lambda^4$ so that the up, down, 
charm, and strange quark Yukawa couplings are of the correct 
order in $\lambda$ (assuming $\tan\beta\sim 1$).  

The lepton Yukawa matrix transforms in the same way as the
down Yukawa matrix transposed.  Therefore, the two-by-two
block of the lepton Yukawa matrix is also determined by the vevs 
of the flavon product $\Phi_Q \Phi_D$.  If this product represented a 
single matrix, then we would obtain the undesirable 
relation $m_e/m_\mu = m_d/m_s$.  However, we have seen that there are 
in fact four contributions to the Yukawa matrices, each multiplied by 
an unknown coefficient of order one.  This gives us enough degrees of 
freedom to suppress the electron mass relative to that of the down quark.
For concreteness, let us assume that $\Phi_Q^{(1)}$ and 
$\Phi_D^{(1)}$ have vevs proportional to $(0,1)$, while $\Phi_Q^{(2)}$ and
$\Phi_D^{(2)}$ have vevs proportional to $(\lambda,\lambda)$.  
If we take the coefficients $c^l_{11}=3$ and $c^l_{22}=1/3$ (where
the $c^l$ are the coefficients for the leptons that are
analogous to the $c^d$ in eq.~(\ref{eq:dimen6})), and take all
other coefficients to be $1$, then we obtain 
$9 m_e/m_\mu = m_d/m_s \sim \lambda^2$, which is an acceptable result.  
Had we required coefficients much larger than $3$ (or much smaller 
than $1/3$), then one might object that the choice of parameters is 
not consistent with naive dimensional analysis.

The remaining diagonal elements of the quark Yukawa matrices consist
of the bottom and top Yukawa couplings.  The bottom Yukawa coupling 
transforms exactly like $\chi_1$, so we require $\delta_1 \sim \lambda^3$.
The top Yukawa coupling is invariant under $(S_3)^3$, and is therefore of 
order $1$ relative to the other elements.

Finally, we must evaluate the other off-diagonal elements of the
up and down Yukawa matrices.  In the down sector, the two-by-one
off-diagonal block transforms as a $({\bf \tilde{2}}, {\bf 1}_A, {\bf 1}_A)
\sim \Phi_Q \chi_2$, and is therefore of the form
\beq
\epsilon_Q \delta_2 \left[\begin{array}{c} \lambda \\ 1
\end{array}\right] \,\,.
\eeq
If we choose $\epsilon_Q \delta_2$ to be of order $\lambda^5$,
then these elements will generate the  Cabibbo-Kobayashi-Maskawa (CKM) 
elements $V_{ub}$ and $V_{cb}$.   The one-by-two block of the down 
Yukawa matrix, which transforms as a $({\bf 1}_S,{\bf 1}_A,{\bf 2})$, is
generated by the product $\Phi_D \chi_1 \chi_2$ and is therefore of the 
form
\beq
\epsilon_D \delta_1 \delta_2 
\left[\begin{array}{cc} \lambda & 1 \end{array} 
\right] \,\,.
\eeq
In the up sector, the off-diagonal block transforming as 
a $({\bf \tilde{2}},{\bf 1}_S,{\bf 1}_S)$ is given by the doublet component
of $(\Phi_Q)^2$.  When taking the product of two
doublets, we will let $\times$ represent the projection
onto the doublet component, $\wedge$ the ${\bf 1}_A$ 
component, and $\cdot$ the ${\bf 1}_S$.  In this case,
we want $\Phi_Q \times \Phi_Q$:
\beq
\epsilon_Q^2 
\left[\begin{array}{c} \lambda \\ 1 \end{array}\right] \,\,.
\eeq
Similarly, the off-diagonal block transforming as a
$({\bf 1}_S, {\bf \tilde{2}}, {\bf 1}_S)$ is given 
by $\Phi_U \times \Phi_U$ and is of the form
\beq
\epsilon_U^2 \, [\lambda \,\,\,\, 1] \,\,.
\eeq

Given the constraints described above 
($\epsilon_Q \epsilon_D\sim\lambda^5$ from the strange mass, 
$\epsilon_Q \epsilon_U\sim\lambda^4$ from the charm mass,
$\delta_1\sim\lambda^3$ from the bottom mass,
and $\epsilon_Q \delta_2 \sim \lambda^5$ to generate adequate
$V_{ub}$ and $V_{cb}$) there is only one set of symmetry breaking
parameters in which no $\epsilon$ or $\delta$ is larger than 
order $\lambda^2$:
\[
\epsilon_Q\sim\lambda^2 \,\,,\,\,\, \epsilon_U\sim\lambda^2
\]\begin{equation}
\epsilon_D\sim\lambda^3 \,\,,\,\,\, 
\delta_1\sim\lambda^3 \,\,,\,\,\, \delta_2\sim\lambda^3
\end{equation}
With this choice, flavor changing neutral current effects 
will not be especially large in any one sector of our model.
Given this choice, we can write down the down and up quark Yukawa 
matrices:
\beq
Y_d \sim \left[ \begin{array}{cc|c} 
                                   \lambda^7    & \lambda^6 & \lambda^6 \\
                                   \lambda^6    & \lambda^5 & \lambda^5 \\
\hline
                                   \lambda^{10} & \lambda^9 & \lambda^3 \\
                \end{array}\right] \,\,,
\eeq
\beq
Y_u \sim \left[\begin{array}{cc|c} \lambda^8 & \lambda^5 & \lambda^5\\
                                  \lambda^7 & \lambda^4 & \lambda^4 \\
\hline
                                  \lambda^5 & \lambda^4 &     1    
\end{array}\right] \,\,.
\eeq
These results are consistent with the masses and mixing angles of 
the Standard Model.

Finally we consider the form of the squark and slepton mass
matrices.  Spurions transforming as either a ${\bf 2}$ or
${\bf 1}_A$ under a single $S_3$ group contribute to the off-diagonal 
entries of the corresponding squark mass matrix.  These 
representations can be formed at lowest order by the products 
$\Phi_a \times \Phi_a$, $\Phi_a^{(1)} \wedge \Phi_a^{(2)} $ or $\Phi_D\chi_2$.  
The analysis is analogous to the one we presented in detail for the quark
Yukawa matrices, so here we will simply quote our results.  The left-handed
squark mass matrices are of the 
form
\beq
m_Q^2 = \left[ \begin{array}{cc|c} 
M_1^2+m^2 \lambda^4 & m^2 \lambda^5 & m^2\lambda^5 \\
m^2 \lambda^5 & M_1^2-m^2 \lambda^4 & m^2\lambda^4 \\
\hline
m^2 \lambda^5 & m^2 \lambda^4 & M_3^2  
\end{array}\right] \,\,.
\eeq
The right-handed squark mass matrices are given by
\beq
m_U^2 = \left[ \begin{array}{cc|c}
M_1^2+m^2 \lambda^4 & m^2 \lambda^5 & m^2\lambda^5 \\
m^2 \lambda^5 & M_1^2-m^2 \lambda^4 & m^2\lambda^4 \\
\hline
m^2 \lambda^5 & m^2 \lambda^4 & M_3^2
\end{array}\right]
\eeq
and
\beq
m_D^2 = \left[ \begin{array}{cc|c}
M_1^2+m^2 \lambda^6 & m^2 \lambda^7 & m^2\lambda^7 \\
m^2 \lambda^7 & M_1^2-m^2 \lambda^6 & m^2\lambda^6 \\
\hline
m^2 \lambda^7 & m^2 \lambda^6 & M_3^2
\end{array}\right] \,\,.
\eeq
All of the off-diagonal elements are consistent with the flavor 
changing neutral current bounds given in Ref.~\cite{MAS}.
The slepton mass matrices $m^2_L$ and $m^2_E$ are of the
same form as $m^2_D$ and $m^2_Q$, respectively.

Finally, we should point out that the supersymmetry breaking
trilinear interactions have the same flavor structure as
the fermion Yukawa matrices, but generally involve different
order one coefficients.  Thus, the trilinear interactions 
are not simultaneously diagonalizable with the Yukawa matrices in general
(unlike the situation in Ref.~\cite{CHM}).  An important constraint 
on the form of these couplings comes from the bounds on $\mu \rightarrow e
\gamma$.  The (12) entry of the left-right slepton mass mixing in 
our model is given by
\begin{equation}
(m^2_{LR})_{21} \sim m_s \lambda A
\end{equation}
This is approximately 20 times larger than the result obtained in
Ref.~\cite{CHM}. If we choose the slepton masses to be of order 300 GeV, the
bino mass and the $A$ parameter to be $\sim 100$ GeV, then our model saturates
the experimental bound Br($\mu\rightarrow e \gamma$) $< 4.9 \times 10^{-11}$. 
Here we use the formulae presented in Ref.~\cite{CHM}. 

\section{The Froggatt-Nielsen Model}
\label{sec:froggatt}

In the previous section we constructed a low-energy effective theory in
which the lowest-dimension nonrenormalizable operators involving the flavon
fields generate acceptable fermion Yukawa matrices when the flavons acquire
vevs, without significantly affecting the degeneracy of the squarks 
(or sleptons) of the first two generations.  If the effective theory
below $M_f$ is generated by integrating out heavy states in a
renormalizable theory, then we will generally obtain some subset of 
the operators described in the previous section.  All operators that are 
consistent with the symmetries of the low-energy theory may not 
necessarily be present.  In building a renormalizable theory of flavor,
we need only to verify that the operators we need for generating 
the elements of the fermion Yukawa matrices are present; our general 
operator analysis tells us {\em a priori} that the full theory will 
otherwise be phenomenologically acceptable.

In this section, we will construct a renormalizable version of our
$(S_3)^3$ model incorporating the mechanism of Froggatt and 
Nielsen \cite{FN}.  We will show that the operators we need to 
account for the fermion masses and mixing angles are generated assuming
that there is a relatively economical set of heavy, vector-like 
particles present at the scale $M_f$.  We will then show that our choice 
of quantum numbers for these fields has an added bonus: all the possible 
renormalizable interactions that violate $R$ parity are forbidden by the 
flavor symmetry.  This implies that no $R$-parity-violating nonrenormalizable 
operators (suppressed by powers of $M_f$ only) are generated when the 
heavy states are integrated out.  While there may be Planck-scale-suppressed
operators that violate $R$ parity and are invariant under the flavor
group, these may be rendered harmless by taking the flavor scale to be 
sufficiently low.   We discuss the implications of this scenario 
at the end of this section.

The flavor quantum number assignments of the vector-like chiral superfields
are given in the first column of Table~\ref{table1}.  The electroweak quantum 
numbers of the heavy, unbarred fields are the same as those of the 
corresponding MSSM field (i.e. $Q^H$ is a color triplet, weak doublet with 
hypercharge $\frac{1}{6}$, etc.)  While we have displayed only one 
generation of the vector-like fields in Table~\ref{table1}, we assume 
the existence of two generations, for reasons detailed below.  In 
addition to the two heavy generations, there are also the 
`extra' heavy fields $L'^H$, $\overline{L'}^H$, $D'^H$, and $\overline{D'}^H$,
also shown in the table.  In SU(5) language, the heavy particle content 
consists of two generations, two antigenerations, and an 
additional {\bf 5}+${\bf \overline{5}}$.  Note that $R$ parity
assignments are also displayed in Table~\ref{table1}.
 
\begin{table}
\begin{center}
\begin{tabular}{cc|cc}
\multicolumn{2}{c}{$R$-parity odd}  & \multicolumn{2}{c}{$R$-parity even} 
\\ \hline
$Q^H \mbox{, } \overline{Q}^H$ & $({\bf 1}_S,{\bf 1}_A,{\bf 1}_S)$  &
$\Phi_Q^{(i)}$  & $({\bf 2},{\bf 1}_A,{\bf 1}_S)$ \\
$U^H \mbox{, } \overline{U}^H$ & $({\bf 1}_A,{\bf 1}_S,{\bf 1}_S)$  &
$\Phi_D^{(i)}$  & $({\bf 1}_A, {\bf 1}_S, {\bf 2})$ \\
$D^H \mbox{, } \overline{D}^H$ & $({\bf 1}_A,{\bf 1}_S,{\bf 1}_S)$  &
$\Phi_U^{(i)}$ & $({\bf 1}_A,{\bf 2},{\bf 1}_S)$ \\
$L^H \mbox{, } \overline{L}^H$ & $({\bf 1}_A,{\bf 1}_S,{\bf 1}_S)$  &
$\chi_1$ & $({\bf 1}_S,{\bf 1}_A,{\bf 1}_A)$ \\
$E^H \mbox{, } \overline{E}^H$ & $({\bf 1}_S,{\bf 1}_A,{\bf 1}_S)$  &
$\chi_2$ & $({\bf 1}_A,{\bf 1}_S,{\bf 1}_A)$  \\
$L'^H \mbox{, } \overline{L'}^H$ & $({\bf 1}_S,{\bf 1}_A,{\bf 1}_S)$ 
&$H_u$ & $({\bf 1}_A,{\bf 1}_A,{\bf 1}_S)$   \\
$D'^H \mbox{, } \overline{D'}^H$ & $({\bf 1}_S,{\bf 1}_A,{\bf 1}_S)$ 
&$H_d$ & $({\bf 1}_A,{\bf 1}_A,{\bf 1}_S)$ \\
\multicolumn{2}{c|}{+ matter} & 
\end{tabular}
\caption{Field content of the theory above the flavor scale.  Only
one generation of the vector-like fields is shown.} 
\label{table1}
\end{center}
\end{table}

Given the particle content in Table~\ref{table1}, it is straightforward
to construct the operators that generate the fermion Yukawa matrices.
Consider the two-by-two block of the down Yukawa matrix.  The relevant
couplings in the superpotential are of the form
\begin{equation}
W= \sum_{ij} (Q \cdot \Phi_Q^{(i)}) \overline{Q}^H_j + Q^H_i H_d D^H_j 
+ (D \cdot \Phi_D^{(i)}) \overline{D}^H_j
\label{eq:rsup}
\end{equation}
where the subscript on the heavy fields indicates the heavy generation
or antigeneration.  By integrating out the heavy 
fields in (\ref{eq:rsup}), we are left with the four operators presented in
equation~(\ref{eq:dimen6}).  This result is represented graphically in 
Figure~1.  Notice that the coupling $Q \Phi_Q^{(i)} \overline{Q}^H$
is involved in generating both the two-by-two up and down
quark Yukawa matrices.  If only one generation of heavy fields were
present, then a single linear combination of $\Phi_Q^{(1)}$ 
and $\Phi_Q^{(2)}$ would enter in these diagrams, and we would be
left with no Cabibbo angle.   We require two heavy generations so that 
two linearly independent combinations of the $\Phi_a^{(i)}$ contribute 
to the operators in the effective theory described in the
previous section.  Note that the couplings $\overline{D}^H_j \chi_2 b$,
$\overline{D'}^H \chi_1 b$, and $Q_3 H_d D'^H$ in the superpotential
are necessary for generating the other elements of $Y_d$.

Notice that the Yukawa matrices are simpler in this model than we would 
have expected from our general operator analysis.  With the particle content 
specified in Table~\ref{table1}, we find that the (3,1) and (3,2) entries 
of the up and down Yukawa matrices as well as the (1,3) and (2,3) entries 
of the up matrix are not generated by heavy particle exchange.  While
sparse, the Yukawa matrices are nonetheless phenomenologically acceptable.

One of the interesting features of the quantum number assignments 
in this model is that it is not possible to write down any 
$R$-parity-violating renormalizable interactions that are invariant under 
the flavor group.  Consider first the $R$-parity-violating operators that 
involve three heavy R-odd fields.  Since each heavy field transforms as 
a ${\bf 1}_A$ under a single $S_3$ group, the product of three can never 
form an invariant.  Next consider the operators that involve two heavy R-odd 
fields and one light matter field.  The product of the two heavy fields 
either forms a singlet or transforms as $({\bf 1}_A,{\bf 1}_A)$ under 
exactly two of the $S_3$ groups.  Since the light field transforms 
nontrivially under a single $S_3$ group, the heavy-heavy-light 
combination can never form an invariant.  The remaining interactions 
involving three R-odd fields are those with zero or one heavy field. 
These are cataloged in Table~\ref{table2}.
\begin{table}
\begin{center}
\begin{tabular}{cc|cc}
Operator &  Transformation & Operator & Transformation \\ \hline
$UDD$ & $({\bf 1}_S,{\bf 2}+{\bf 1}_A,{\bf 1}_A)$ & 
$L^HLE$ & $({\bf 2}+{\bf 1}_S, {\bf 1}_S, {\bf 2}+{\bf 1}_A)$ \\
$QLD$ & $({\bf 2}+{\bf 1}_A, {\bf 1}_S, {\bf 2}+{\bf 1}_A+{\bf 1}_S)$ &
$L'^HLE$ & $({\bf 2}+{\bf 1}_A,{\bf 1}_A,{\bf 2}+{\bf 1}_A)$ \\
$LLE$ & $({\bf 2}+{\bf 1}_A, {\bf 1}_S, {\bf 1}_A)$ & 
$LLE^H$ & $({\bf 1}_S, {\bf 1}_A, {\bf 1}_A)$ \\
$U^HDD$ & $({\bf 1}_A, {\bf 1}_S, {\bf 1}_A)$ & 
$QQ\overline{D}^H$ & $({\bf 2}+{\bf 1}_A,{\bf 1}_S,{\bf 1}_S)$ \\
$UD^HD$ & $({\bf 1}_A, {\bf 2}+{\bf 1}_A, {\bf 2}+{\bf 1}_A)$
&$QQ\overline{D}'^H$ & $({\bf 2}+{\bf 1}_S,{\bf 1}_A,{\bf 1}_S)$\\
$UD'^HD$ & $({\bf 1}_S, {\bf 2}+{\bf 1}_S, {\bf 2}+{\bf 1}_A)$
&$Q\overline{L}^HU$ & $({\bf 2}+{\bf 1}_S,{\bf 2}+{\bf 1}_A,{\bf 1}_S)$\\
$Q^HLD$ & $({\bf 1}_S, {\bf 1}_A, {\bf 2}+{\bf 1}_A+{\bf 1}_S)$
&$Q\overline{L'}^HU$&$({\bf 2}+{\bf 1}_A,{\bf 2}+{\bf 1}_S,{\bf 1}_S)$\\
$QL^HD$ & $({\bf 2}+{\bf 1}_S, {\bf 1}_S, {\bf 2}+{\bf 1}_A)$ &
$U\overline{D}^HE$ & $({\bf 2}+{\bf 1}_S,{\bf 2}+{\bf 1}_A,{\bf 1}_S)$\\
$QL'^HD$ & $({\bf 2}+{\bf 1}_A, {\bf 1}_A, {\bf 2}+{\bf 1}_A)$
&$U\overline{D'}^HE$ & $({\bf 2}+{\bf 1}_A,{\bf 2}+{\bf 1}_S,{\bf 1}_S)$\\
$QLD^H$ & $({\bf 2}+{\bf 1}_S, {\bf 1}_S, {\bf 2}+{\bf 1}_A)$  & & \\
$QLD'^H$ & $({\bf 2}+{\bf 1}_A, {\bf 1}_A, {\bf 2}+{\bf 1}_A)$ & &      
\end{tabular}
\caption{Trilinear operators involving three R-odd fields, with 
zero or one heavy field.} 
\label{table2}
\end{center}
\end{table}

In almost every interaction shown in Table~\ref{table2}, at least one of 
the three fields involved transforms under a different $S_3$ group than
that of the other two, so that there is no possibility of forming an 
invariant.  The only exception is the operator $QQ\overline{D}^H$, which 
involves three fields that each transform under $S_3^Q$.  In this case, 
however, the operator is symmetric under interchange of the two $Q$ 
fields, so we can never form the ${\bf 1}_A$ that we would 
need to produce an invariant.  

The remaining trilinear operators that we need to consider are those 
that involve one R-odd and two even fields.  Since the R-odd fields
all carry electroweak quantum numbers, these operators must be
of the following form to preserve electroweak gauge invariance:
$LH_dF$, $L^H H_d F$, $L'^H H_d F$, $\overline{L}^H H_u F$ or 
$\overline{L'}^H H_u F$, where $F$ is a flavon field (either
$\Phi$ or $\chi$).  The product of the first two fields in each of 
these interactions transform  as a $({\bf 1}_A,{\bf 1}_A,{\bf 2}+{\bf 1}_A)$,
$({\bf 1}_S,{\bf 1}_A,{\bf 1}_S)$, $({\bf 1}_A,{\bf 1}_S,{\bf 1}_S)$ $({\bf
1}_S,{\bf 1}_A,{\bf 1}_S)$ and $({\bf 1}_A,{\bf 1}_S,{\bf 1}_S)$ respectively.
Since the flavon fields transform under exactly two $S_3$ groups, while 
the representations above involve either one or three $S_3$ groups, no 
invariants are possible.  As a corollary, we have shown that all 
the dimension-2 R-odd operators in the superpotential transform nontrivially 
under the flavor group, and are forbidden as well.

$R$ parity is an accidental symmetry in our $(S_3)^3$ model, a consequence of
both the flavor symmetry and the particle content given in Table~1.  Our
preceding discussion, however, has two limitations.  First, we may need to
enlarge the particle content of the model to construct a renormalizable
potential for the flavon fields that yields the pattern of expectation values
assumed in Section~2.  We show in the Appendix that the additional fields
required to construct a suitable potential do not have interactions that spoil
the accidental $R$ parity described in this section.  Secondly, we
have restricted ourselves to a renormalizable Lagrangian.  There may be
non-renormalizable interactions induced at the Planck scale, and some of
these may violate $R$ parity.  Of course, Planck-suppressed $R$-parity
violating operators simply may not be present; it is known, for example,
that superstring compactification usually does not lead to the most
general Lagrangian consistent with the symmetries of the low-energy
theory.  However, it is interesting to consider the constraints on our
model if such $R$-parity-violating operators are indeed generated at the
Planck scale.

The most stringent constraint on $R$-parity violation comes from
non-observation of nucleon decay.  The most dangerous combination of 
operators is $uds$ and $Q_1 s L_{1,2}$, where the subscript is the
generation index.  Since we must combine each of these
with at least two flavon fields to form an $(S_3)^3$ invariant at
the Planck scale, both trilinears are suppressed by $(M_f/M_*)^2$ in 
the low-energy theory, where  $M_*=M_{Pl}/\sqrt{8\pi}$ is the reduced 
Planck mass.  There are operators involving third generation 
fields and/or heavy Froggatt--Nielsen fields, however, that can be constructed
using only one flavon field, yielding trilinear operators that are suppressed
by one power of $(M_f/M_*)$.  Since the third generation and
the heavy fields mix with the first generation fields, dangerous 
operators may result \cite{AG,SV}.  There are two
$UDD$-type operators allowed at linear order in the flavor symmetry
breaking and also linear order in either third generation or heavy fields:
$\chi_1 U_3 (D \wedge D)/M_*$ and $\chi_2 U^H (D \wedge D)/M_*$.  
Given the structure of the Yukawa matrices, $U_3$ does not mix
with the first generation fields (recall that the (3,1) and (3,2)
entries of $Y_u$ were not generated in the full theory) while $U^H$ mixes
at order $\epsilon_U \lambda^3 \simeq \lambda^5$.  Similarly, there are 
three $QDL$-type operators at linear order in spurion and also linear in
either third generation or heavy fields: $\chi_2 Q_3 (D \wedge L)/M_*$, $Q_3
(\Phi_D\cdot (D \times L))/M_*$ and $\chi_1 Q^H (D \wedge L)/M_*$.  The last 
one dominates among these three.  Assuming that these operators are present,
they are tightly constrained from proton decay \cite{HK}:
\begin{equation}
\frac{\delta_2 \epsilon_U \lambda^3 M_f}{M_*}
      \frac{\delta_1 \epsilon_Q \lambda M_f}{M_*}
      \lesssim 10^{-24} \, .
\end{equation}
With our previous choice 
$\epsilon_U \simeq \epsilon_Q \simeq \lambda^2$ and $\delta_1
\simeq \delta_2 \simeq \lambda^3$, we obtain an upper bound on the
flavor scale
\begin{equation}
M_f \lesssim 8 \times 10^{10}~\mbox{GeV} \, .      \label{Mfbound}
\end{equation}
Given this bound, the coefficients $h$ of the $R$-parity-violating 
operators are always smaller than 
$\lambda^2 M_f/M_* \lesssim 2 \times 10^{-9}$,
and all existing experimental bounds are satisfied (for a comprehensive
discussion of these bounds, see {\it e.g.}\/, Refs.~\cite{BGH} or
\cite{recent,AG}); the tightest bound on the $h$ comes from $n$-$\bar{n}$
oscillation with $h \lesssim 10^{-7}$.  Note that the bound from sphaleron
erasure of the cosmic baryon asymmetry $h \lesssim 10^{-8}$ \cite{CDEO} is 
also satisfied.\setcounter{footnote}{0}~\footnote{This bound may be even
weaker in some cases
\cite{DR}.}

There is a potentially strong constraint from cosmology if the
$R$-parity violation is very weak. The lightest neutralino may 
decay after big bang nucleosynthesis and spoil its successful
predictions \cite{KaM}.  For instance, we can estimate the lifetime of 
a bino-like neutralino assuming it decays via squark exchange and 
an $R$-parity-violating trilinear coupling:
\begin{equation}
\Gamma_{\tilde{\chi}_1^0} \sim \frac{1}{64 \pi^2}
      \frac{\alpha}{\cos^2 \theta_W} 
      \left(\frac{h}{m_{\tilde{q}}^2} \right)^2
      m_{\tilde{\chi}_1^0}^5 \, .
\end{equation}
If we take $h = \lambda^2 M_f/M_*$, 
$m_{\tilde{\chi}_1^0} \sim 100$~GeV, $m_{\tilde{q}} \sim
1$~TeV, and $M_f \sim 10^{10}$~GeV, we obtain the lifetime
$\tau_{\tilde{\chi}_1^0} \sim 20$~sec.  This satisfies the constraint
from nucleosynthesis on a long-lived particle decaying into jets $\tau
\lesssim 10^3$~sec \cite{Sarkar}.  The constraint is weaker
($\tau \lesssim 10^6$~sec) if $\tilde{\chi}_1^0$ decays primarily into
photons or leptons \cite{Moroi}.\footnote{If the neutralino is too
abundant, corresponding to $\Omega_\chi \gtrsim 10^2$ in the stable limit,
and has a lifetime longer than 1~sec, it contributes to the energy density
of the Universe and affects the expansion rate when the neutron abundance
freezes out, and spoils the standard big bang nucleosynthesis
predictions.  Recall, however, the neutralino abundance is typically
between $\Omega_\chi \sim 10^{-3}$ to $10^2$.}  The constraint from the
distortion in the cosmic microwave background spectrum is weaker than
the one from nucleosynthesis \cite{HUS}.

For completeness, it is important to consider the proton decay 
constraints on Planck-suppressed dimension-five operators as well.
Recall that in Ref.~\cite{CHM}, we used these bounds to restrict
the transformation properties of the flavon fields, assuming that
the flavor scale was identical to the Planck scale.  However, when
$M_f < M_*$, the dimension-five operators are significantly 
suppressed. The largest dimension-five operators in our model
are generated from the following flavor-invariant dimension-6 
operators: $(Q\cdot Q)(Q_3 \Phi_D \cdot L) / M_*^2$ and
$(Q \cdot Q) (Q_3 \chi_2 L_3) / M_*^2$.  When the flavon fields
acquire vevs, these operators generate dimension-five operators
with coefficients $(M_f/M_*) (\lambda^3/M_*)$.  The third generation 
doublet field mixes with the second generation at order $\lambda^2$.
Thus, the coefficient of the operator that directly contributes
to the decay is $(M_f/M_*)(\lambda^5/M_*)$.  If we compare this
to the experimental bound, which requires the coefficient to be
smaller than ${\cal O}(\lambda^8/M_*)$ \cite{CHM}, then we obtain 
\begin{equation}
M_f \lesssim 10^{16} \mbox{ GeV}
\end{equation}
This bound is much weaker than the one we obtained from the 
$R$-parity-violating operators in eq.~(\ref{Mfbound}).

Finally, we should mention that the gauge coupling constants become
non-perturbative below the Planck scale in our model, assuming that 
the vector-like particles are integrated out at a scale $M_f$
satisfying Eq.~(\ref{Mfbound}).  If we require perturbativity of
the gauge couplings up to the scale $M_*$, then we obtain the
lower bound $M_f \gtrsim 3 \times 10^{12}$~GeV.  However, we do not
consider this as a serious problem of the model since this scale is rather
close to the upper bound given in Eq.~(\ref{Mfbound}).  The particle content
or gauge group may be altered close to the Planck scale, or one may go 
over to the dual description of the theory which remains weakly coupled.

\section{Conclusions}

We have presented a supersymmetric theory of flavor and $R$ parity
based on the discrete flavor group $(S_3)^3$.  After specifying
the flavor symmetry breaking fields, we showed that the most general 
low-energy effective theory consistent with the flavor and gauge 
symmetries does not lead to large flavor changing neutral 
current effects.  The hierarchical pattern of the fermion Yukawa 
matrices and the near degeneracy of the squarks (or sleptons) of the 
first two generations are both guaranteed in our model by the 
flavor symmetry.  In addition, we showed that an acceptable effective 
theory could originate from a renormalizable model via the 
Froggatt-Nielsen mechanism, and we presented an economical set of 
heavy vector-like fields responsible for generating the necessary operators.  
After specifying the particle content of the theory above the flavor 
scale $M_f$, we showed that all renormalizable operators that violate 
$R$ parity were forbidden by the flavor symmetry.  Thus, at the 
renormalizable level, $R$ parity arose as an accidental symmetry in 
our model, a consequence of the flavor group and particle content.  
Furthermore, we showed that $R$-parity-violating nonrenormalizable 
operators generated at the Planck scale could be sufficiently suppressed 
by taking the flavor scale to be less than $10^{11}$ GeV.  Our model 
demonstrates that it is possible to explain simultaneously the hierarchical 
form of the fermion Yukawa matrices, the suppression of flavor changing 
neutral current processes, and the absence of renormalizable baryon and 
lepton number violating couplings in supersymmetric  
models by introducing a flavor group and a specific mechanism of
flavor symmetry breaking.

In section 2 we stressed that supersymmetric theories require some new
symmetry, which we called $X$, to suppress $B$ and $L$ violation, and that
there are many candidates for $X$. It is interesting to compare the $X$
symmetry introduced in this paper with other elegant possibilities.

It is possible for $X$ to be a discrete gauge symmetry, the most compelling of
which is the $Z_2$ subgroup of $SO(10)$ generated by the element
$$
X(SO(10)) = e^{i \pi (2T_{3_L} + 2 T_{3_R})} = e^{i \pi N_s}
\eqno(I)
$$
where $N_s$ is 1 for spinorial representations and zero otherwise.
When the rank of $SO(10)$ is broken, a special choice of representation or
further discrete symmetry  is required to ensure that this $X$ symmetry is
left unbroken.

An elegant flavor group origin for $X$ is possible with a flavor group $U(3)$,
which contains a $Z_2$ with element
$$
X(U(3)) =  e^{i \pi N_T} \eqno(II)
$$
where $N_T$ is the triality of the representation. $X$ conservation of the
low energy theory follows if all flavor violation, in particular that which
generates the quark and lepton masses, is generated by vevs of flavon fields
with $N_T$ even. 

In the $(S_3)^3$ model of this paper, the $X$ symmetry can similarly be
defined
as a $Z_2$ generated by an element which depends on representation type:
$$
X(S_3^3) =  e^{i \pi (N_{1_A} + N_2)} \eqno(III)
$$
where $N_{1_A}, N_2$ count the number of ${\bf 1}_A, {\bf 2}$ representations
of a given field. (For example, the representation 
$({\bf 2},{\bf 1}_A,{\bf 1}_S)$ has $N_{1_A}+N_2=2$.)
This $X$ will not be spontaneously broken  if all Higgs and
flavon fields have $N_{1_A} + N_2$ even, as occurs in the model of this 
paper\footnote{To forbid all the phenomenologically dangerous operators,
it is necessary only for $X$ to be a symmetry of the matter fields.}.

From equations (I,II,III), one sees that these three examples of $X$ symmetry
have a comparable elegance. However, there is an important distinction. In
cases I,II the symmetry group SO(10), U(3) is sufficient to ensure that $X$
is an exact symmetry of the Lagrangian; indeed, $X$ is a discrete
subgroup of the gauge or flavor symmetry. This is not true in the case III:
$X$ is explicitly broken by any ${\bf 2}^3$ or ${\bf 2}^2 {\bf 1}_A$ invariant
allowed by the gauge symmetry. Hence in case III, explicit violations of $B$
and $L$ are expected at some level, and the LSP is not expected to be
absolutely stable.

\begin{center}               
{\bf Acknowledgments} 
\end{center}
We thank Nima Arkani-Hamed, Hsin-Chia Cheng, and Takeo Moroi for useful 
comments.
This work was supported in part by the Director, Office of 
Energy Research, Office of High Energy and Nuclear Physics, Division of 
High Energy Physics of the U.S. Department of Energy under Contract 
DE-AC03-76SF00098 and in part by the National Science Foundation under 
grant PHY-90-21139.

\appendix

\section{Flavon Potential}

In this appendix we present a possible form of the potential for the
flavon fields.  We discuss this issue for the following reasons.
First, it is not possible to generate flavon vevs via
a renormalizable potential using the flavon fields presented in the 
main body of the paper alone.  If we rely only on the
minimal flavon content, we must rely on higher dimension operators
to obtain the desired form of the expectation values.
If the higher dimension operators arise at the Planck scale, we
obtain typical flavon masses of order $m_\phi \sim (\lambda^2 M_f)^2
/ M_*$.  Furthermore, if we require that $M_f$ satisfy the upper bound given 
in Eq.~(\ref{Mfbound}), then the flavon fields turn out to be rather light,
$m_\phi \lesssim 400$~MeV.  Unless one arranges the scales such that
$m_\phi>m_K-m_\pi$, we will have the dangerous flavor-changing decays $K^+
\rightarrow \pi^+ \phi$ or $\mu^- \rightarrow e^- \phi$ at rates beyond 
the experimental bounds.\footnote{For instance, the effective operator
generated by Froggatt--Nielsen fields $W = (H_Q Q) (H_D D) H_d/M_f^2$ 
gives us an operator 
$W = (\epsilon_Q \langle H_d \rangle / M_f) \overline{d_R} s_L \varphi
$, where $\varphi$ is the physical field corresponding to the upper
component of $H_D$. On the other hand, $K^+ \rightarrow \pi^+ \phi$ with a
massless $\phi$ constrains the coupling $(1/F) \partial_\mu \phi \bar{d}
\gamma^\mu s$ such that $F \gtrsim 10^{11}$~GeV. If $\varphi$ is light, 
we obtain $M_f \gtrsim 10^{13}$~GeV.} The simplest way to avoid this potential 
phenomenological disaster is to arrange for renormalizable couplings 
among the flavon fields themselves to generate flavon masses of order $M_f$.
Second, if we extend the particle content of flavons in a way that
allows us to write down an explicit renormalizable potential, we may find
that $R$ parity is no longer an accidental consequence of the flavor
symmetry and particle content, as emphasized in Section~3.  The danger is 
that the new flavons may couple directly to the ordinary matter
fields, and generate flavor-invariant, renormalizable $R$-odd couplings.
The purpose of this section is to show that an extension of the particle content
that allows us to write down a suitable potential for the flavon fields 
still preserves the accidental $R$ parity of the minimal theory.

Writing down a potential for $\chi_{1,2}$ fields is easy.  One needs to
introduce fields $\xi$ which transforms as a $({\bf 1}_S, {\bf 1}_S, {\bf
1}_S)$. The most general renormalizable potential is then
\begin{equation}
W = \frac{1}{2} m_\chi \chi^2 + \frac{1}{2} m_\xi \xi^2 
      - g_\chi \chi^2 \xi - g_\xi \xi^3 \, .
\end{equation}
This potential has a stationary configuration,
\begin{eqnarray}
\xi &=& \frac{m_\chi}{2 g_\chi} \, , \\
\chi &=& \sqrt{(m_\xi \xi + 3 g_\xi \xi^2)/g_\chi} \, .
\end{eqnarray}
Since $\xi$ does not carry any flavor quantum number, none of our
previous conclusions are affected by its existence.

Constructing a potential for $\Phi_{Q,U,D}$ is slightly more difficult.  
Since all $\Phi$'s have one doublet and one ${\bf 1}_A$ factor, different
types of $\Phi$'s cannot couple to each other in the renormalizable
superpotential.  Therefore, we consider potentials for different types
of $\Phi$'s separately and discuss a $\Phi$ field generically transforming 
as a $({\bf 2}, {\bf 1}_A)$ under $(S_3)^2$ without worrying which two
$S_3$ groups are involved.   Let us introduce another doublet 
field $K \sim ({\bf 2}, {\bf 1}_S)$.  The most general
renormalizable potential is\footnote{There may be couplings of the type
$\Phi^2 \xi$ or $K^2 \xi$.  However, these coupling do not affect the
stationary configurations we discuss, and can be absorbed into $m_\Phi$ and
$m_K$ by a redefinition.}
\begin{equation}
W = \frac{1}{2} m_\Phi \Phi^2 + \frac{1}{2} m_K K^2 
      - g_\Phi (\Phi \times \Phi) \cdot K - g_K (K \times K) \cdot K .
\label{eq:ppot}
\end{equation}
The reader should not worry that the third and fourth terms are
$X$-violating couplings.  Since $K$ does not couple directly to any
of the fields in the first column of Table~\ref{table1}, $X$ remains 
conserved on the matter fields.  This potential (\ref{eq:ppot}) allows a
stationary configuration
\begin{eqnarray}
\Phi &=& \left( \begin{array}{c}
            0 \\ \sqrt{(m_K K_1 + 3 g_K K_1^2)/g_\Phi}
           \end{array} \right), \\
K &=& \left( \begin{array}{c}
            m_\Phi/2 g_\Phi \\ 0
          \end{array} \right).
\end{eqnarray}
Note that this configuration leaves a non-trivial $S_3$ subgroup
unbroken 
\[
S_3 = \{ (e,e),\ (e,(123)),\  (e,(132)),\  ((12),(12)),\  
((12),(23)),\  ((12),(31))\} \,
\]
and hence the existence of this extremum
is guaranteed by the symmetry.  By having another independent set of
$\Phi'$ and $K'$, one may have the same type of extremum but with a $Z_3$
rotation,
\begin{eqnarray}
\Phi' &=& \left( \begin{array}{cc}
      -1/2 & \sqrt{3}/2 \\
      -\sqrt{3}/2 & -1/2
           \end{array} \right) 
      \left( \begin{array}{c}
            0 \\ \sqrt{(m'_K K'_1 + 3 g'_K K_1^{\prime 2})/g'_\Phi}
           \end{array} \right) \\
K' &=& \left( \begin{array}{cc}
      -1/2 & \sqrt{3}/2 \\
      -\sqrt{3}/2 & -1/2
           \end{array} \right) 
      \left( \begin{array}{c}
            m'_\Phi/2 g'_\Phi \\ 0
          \end{array} \right).
\end{eqnarray}
If the overall scale of $\Phi'$, $K'$ is lower than $\Phi$ and $K$ by a factor
of $\lambda$, we obtain the desired form of the expectation values of
$\Phi$ and $\Phi'$.
\footnote{If a coupling between $\Phi$, $K$ sector and $\Phi'$,
$K'$ sector is present, such as $(\Phi \times \Phi') \cdot K$, the minima are
shifted due to mixing between $\Phi$ and $\Phi'$.  Such a mixing makes both
components of $\Phi$ and $\Phi'$ non-vanishing, and does not lead to any
problem.} 

The important point is that $K$ fields do not contribute to the mixing
between light and Froggatt--Nielsen fields because they lack the ${\bf
1}_A$ factor.  It is easy to check that none of our conclusions
regarding the form of the Yukawa matrices, scalar matrices, and the 
accidental $R$ parity present at the renormalizable level are modified 
by the existence of the $K$ fields.  Our discussion of nonrenormalizable
$R$-parity-violating operators is only slightly modified, by the
existence of the operator $W = (K_Q \cdot Q) (d \cdot L)/M_*$.  If the
expectation value of $K_Q$ is similar to that of $\Phi_Q$, this operator
gives an $R$-parity violating $Q_1 s L_2$ operator with a coupling of
$\epsilon_Q \lambda M_f/M_*$, which is larger than that discussed in
section~3 by $\lambda^3$.  The upper bound on $M_f$ in
Eq.~(\ref{Mfbound}) is strengthened by $\lambda^{3/2}$, or $M_f \lesssim 8
\times 10^9$~GeV.  Note, however, that the expectation value
of $K$ can be made different from $\Phi$ by varying $m_K$ from $m_\Phi$. 
Hence the bound given in Eq.~(\ref{Mfbound}) is the only one that is
parameter-independent.



\begin{center}
{\bf Figure Captions}
\end{center}

{\bf Fig. 1}   Diagrammatic representation of the operators generated by 
heavy particle exchange.  The operators shown contribute to the up
and down quark Yukawa matrices when the flavons acquire vacuum expectation
values.
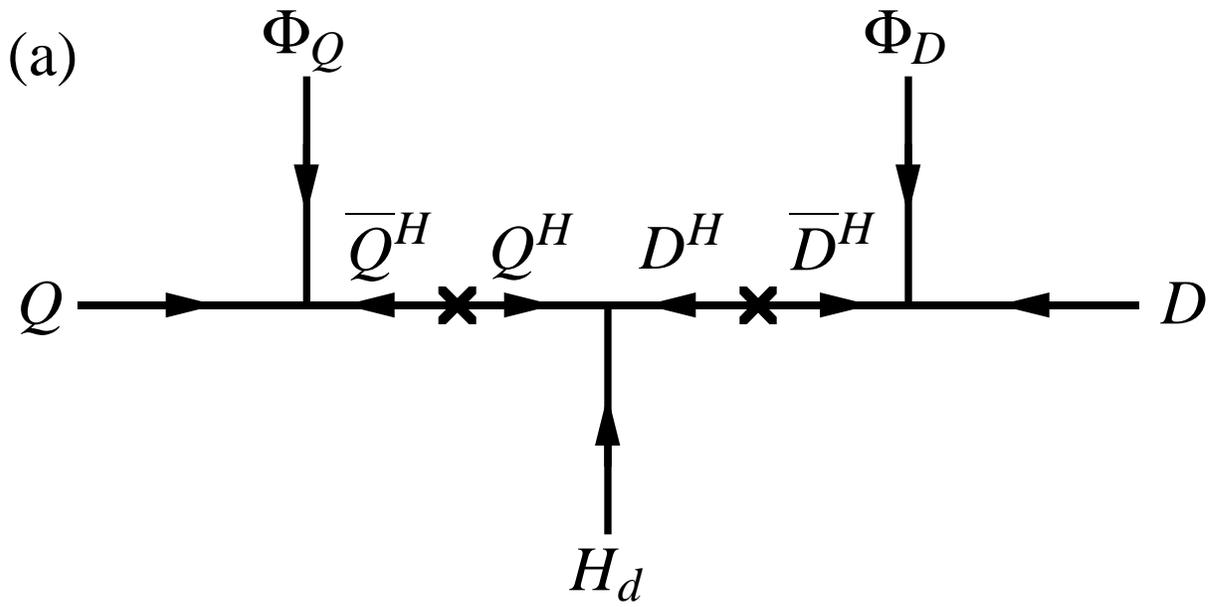
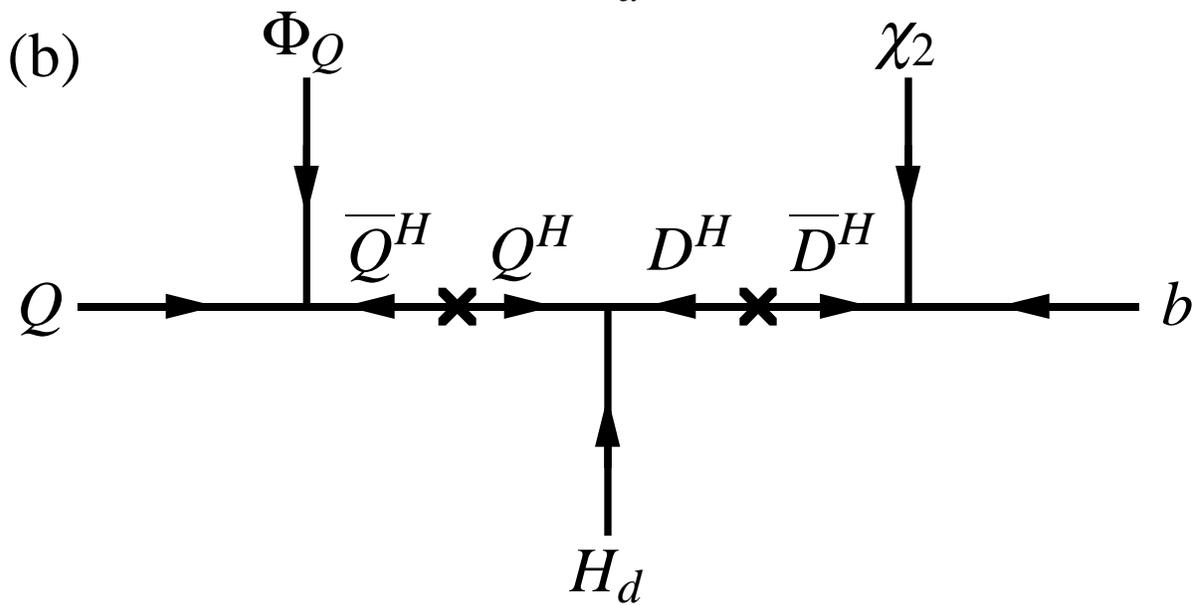
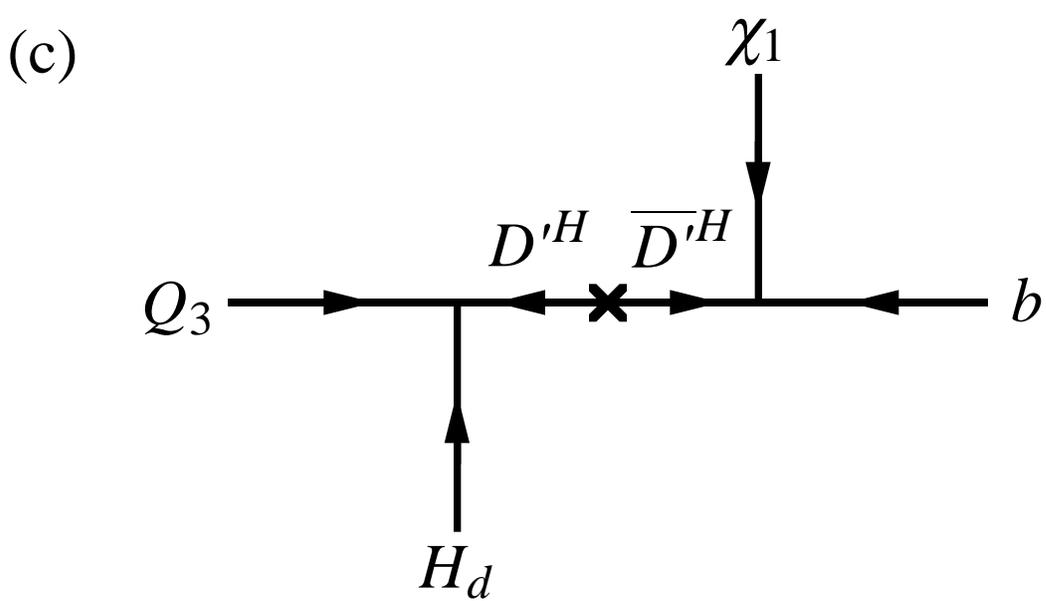
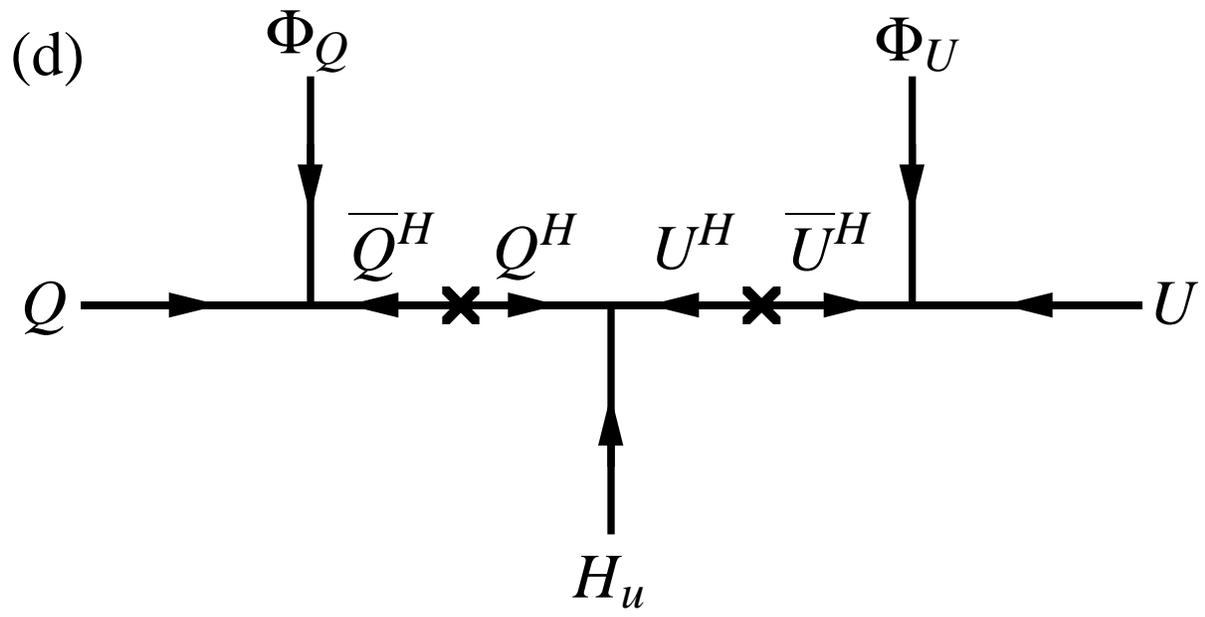
\end{document}